\documentclass[groupedaddress,nofootinbib,showpacs,eqsecnum]{revtex4}

\usepackage{euscript,graphicx,amssymb,subfigure}

\usepackage{amsfonts}
\usepackage{amsmath}
\usepackage{amssymb}
\usepackage{graphicx}
\usepackage{euscript}
\usepackage{color}
\usepackage{subfigure}

\begin{document}
\title{Interacting agegraphic dark energy model in tachyon cosmology coupled to matter  }

\author{H. Farajollahi}
\email{hosseinf@guilan.ac.ir} \affiliation{Department of Physics, University of Guilan, Rasht, Iran}
\affiliation{School of Physics, University of New South Wales, Sydney, NSW, 2052, Australia}
\author{A. Ravanpak}
\email{aravanpak@guilan.ac.ir} \affiliation{Department of Physics,
University of Guilan, Rasht, Iran}
\author{G. F. Fadakar}
\email{gfadakar@guilan.ac.ir} \affiliation{Department of Physics,
University of Guilan, Rasht, Iran}

\date{\small {\today}}

\def\be{\begin{equation}}
\def\ee{\end{equation}}
\def\bea{\begin{eqnarray}}
\def\eea{\end{eqnarray}}
\def\M{{{\cal M}}}
\def\bdy{{\partial\cal M}}
\def\w{\widehat}
\def\n{\widetilde}
\def\real{{\bf R}}

\begin{abstract}
Scalar-field dark energy models for tachyon fields are often regarded as an effective description of
an underlying theory of dark energy. In this paper, we propose the agegraphic
dark energy model in tachyon cosmology by interaction between the components of the dark sectors. In the formalism, the interaction term emerges from the tachyon field nonminimally coupled to the matter Lagrangian in the model rather than being inserted into the formalism as an external source. The model is constrained by the observational data. Based on the best fitted parameters in both original and new agegraphic
dark energy scenarios, the model is tested by Sne Ia data. The tachyon potential and tachyon field are reconstructed and coincidence problem is revisited.
\end{abstract}

\maketitle

\newpage

\section{Introduction}

The current cosmic acceleration is the subject of both theoretical and observational cosmology in recent years \cite{Reiss}. A component which causes cosmic
acceleration is usually dubbed dark energy (DE) which is part of a mysterious puzzle in modern cosmology.
Among all, the cosmological constant introduced by Einstein was the first serious challenging candidate for DE with the so called fine tuning and cosmic coincidence issues \cite{Weinberg}. Alternatively, the holographic dark energy (HDE)
and agegraphic dark energy (ADE), both appear to be consistent with quantum kinematics, in the sense that obey the
Heisenberg type uncertainty relation and predict a time-varying DE equation of state (EoS). In HDE models the event horizon plays the role of cosmological length scale. These models, that has recently been studied
widely, are very successful in explaining the observational data \cite{Set}--\cite{Cohen1}. However, an obvious drawback concerning
causality appears in these models by choosing event horizon as the length scale.
More recently, a new DE model, dubbed "agegraphic
dark energy" model, has been proposed by Cai \cite{Cai}, which is also related to the holographic
principle of quantum gravity. The model reconsider both the uncertainty
principle in quantum physics and the gravity in general relativity \cite{jamil}-\cite{karami}.
The ADE model which has been tested by observation \cite{skeyhki3}, expels the causality problem in HDE, since the length scale is just the age of the  universe. One crucial advantage of HDE and ADE over other DE models is that they initiated from quantum physics \cite{Sen}.

For the first time, the authors in \cite{wang} studies the interaction between HDE and DM in IHDE models, based on phenomenological grounds, with the aim to alleviate the coincidence problem.

In general, the interacting terms in IADE models are not unique. Here, we would like to extend the previous work carried in the IADE models, by studying a tachyon cosmological model in which the scalar field in the formalism plays two roles: as a scalar field interact with the matter in the universe and as a tachyon field plays the role of DE. In our formalism the interacting term naturally appears in the model from the interaction between scalar field and matter field in the universe.
We consider the cosmological model in the presence of  with a tachyon potential and a nonminimally scalar field coupled to the matter lagranigian  in the action given by \cite{faraj}-\cite{faraj1},
\begin{eqnarray}\label{action}
S=\int[\frac{M_p^2R}{2}-V(\phi)\sqrt{1-\partial_\mu \phi \partial^\mu \phi}+f(\phi)\mathcal{L}_{m}]\sqrt{-g}d^{4}x,
\end{eqnarray}
where $R$ is Ricci scalar. Unlike the usual Einstein-Hilbert action, the matter
Lagrangian ${\cal L}_{m}$ is modified as $f(\phi){\cal L}_{m}$, where $f(\phi)$ is
an analytic function of $\phi$. This last term in Lagrangian brings about the nonminimal
interaction between the matter and the scalar field. The investigations on the reconstruction of the tachyon potential $V (\phi)$
in the framework of ADE have been carried out in \cite{setare}. One of the motivation to include the interaction between agegraphic description of tachyon DE and CDM is to solve the coincidence problem \cite{Baglaa}--\cite{Baglaa1}. In the next two sections we reconstruct the potential and the dynamics of the tachyon scalar field in both original and new ADE.

\section{TACHYON RECONSTRUCTION OF THE ORIGINAL ADE}

The variation of action (\ref{action}) with respect to the metric tensor components in a spatially flat FRW  cosmology yields the following field equations,
\begin{eqnarray}
3H^{2}M_p^2&=&\rho_{m}f+\frac{V(\phi)}{\sqrt{1-\dot{\phi}^{2}}},\label{fried1}\\
M_p^2(2\dot{H}&+&3H^2)=-\gamma\rho_{m}f+V(\phi)\sqrt{1-\dot{\phi}^{2}},\label{fried2}
\end{eqnarray}
where $ H=\frac{\dot{a}}{a}$ is the Hubble parameter. In the above, we also assumed a perfect fluid filled the universe with the equation of state $p_{m}=\gamma\rho_{m}$. In the following we assume that the matter in the universe is CMD where $\gamma=0$ . We can rewrite the above equations as
\begin{eqnarray}
3H^{2}M_p^2&=&\rho_{m}f+\rho_{tac},\label{fried11}\\
M_p^2(2\dot{H}&+&3H^2)=-p_mf-p_{tac},\label{fried22}
\end{eqnarray}
where $\rho_{tac}$ and $p_{tac}$ are respectively the energy density and pressure of the tachyon field. We define the fractional energy densities such as
\begin{eqnarray}
\Omega_{mf}=\frac{\rho_{m}f}{3M_p^2H^2} & , &\Omega_{tac}=\frac{\rho_{tac}}{3M_p^2H^2},\label{o}
\end{eqnarray}
where ``$mf$'' and ``$tac$'' stand for nonminimally coupled scalar field to matter lagranigian and tachyon, where $\Omega_{mf}=\Omega_mf$. Thus, the Friedmann equation can be written as
\begin{eqnarray}\label{cons}
\Omega_{mf}+\Omega_{tac}&=&1,\label{conserv}
\end{eqnarray}
Next we intend to implement the interacting original ADE models with tachyon scalar field.
Let us first review the origin of the ADE model. From quantum fluctuations of
spacetime \cite{Karolyhazy}, the time parameter $t$ in Minkowski spacetime in not more accurate than $\delta t = \beta t_p^{2/3}t^{1/3}$ where $\beta$ is a dimensionless constant of order unity. As a result, the energy density is given by \cite{m}-\cite{li}
\begin{eqnarray}
\rho_{D}&\sim&\frac{1}{t_p^2t^2}\sim\frac{M_p^2}{t^2},\label{rho}
\end{eqnarray}
where $t_p$ is the reduced Planck time. In \cite{Cai}, the DE density in given by (\ref{rho}) with $t$ as the universe age:
\begin{equation}\label{T}
T=\int_0^a\frac{da}{Ha}.
\end{equation}
From (\ref{rho}) and (\ref{T}), the original AD energy density is
\begin{equation}\label{rhoD}
\rho_D=\frac{3n^2M_p^2}{T^2}.
\end{equation}
The factor $3n^2$ parameterizes some uncertainties in the model, such as the
species of quantum fields in the universe, the effect of curved space-time, and so on. The relation (\ref{rhoD}) is similar to what we have in  HDE, except the length measure that is the age of the universe. From (\ref{rhoD}) and  (\ref{o}) , we obtain
\begin{equation}\label{omega}
\Omega_D=\frac{n^2}{T^2H^2}.
\end{equation}
If we assume that the scalar field as a tachyon field play the role of ADE and as a nonminimally coupled field play the role of DM, then with the interaction between these two fields their energy densities no longer satisfy independent conservation
laws, instead they obey:
\begin{eqnarray}\label{ct1}
\dot{\rho_{mf}}&+&3H\rho_{mf}=Q,
\end{eqnarray}
\begin{equation}\label{ct}
\dot{\rho_{tac}} + 3H(1+\omega_{tac})\rho_{tac}=-Q,
\end{equation}
 where $\rho_{mf}=\rho_m f$ and $Q = \rho_{m}\dot{f}$ is the interaction term. In $Q$, $\dot{f}$ gauges the intensity of the coupling between matter and scalar field. For $\dot{f}=0$, there is no interaction between DM and ADE.
 The $Q$  term measures the different evolution of the DM due to its interaction with the ADE which gives rise to a
different universe expansion. The interesting point concerning the interaction term is that in comparison to the other agegraphic models where the form of the interaction term $Q$ is not unique and usually is expressed as $Q=3b^2H(\rho_m+\rho_{DE})$, in our model the interaction term naturally appears in the model directly as a function of the scalar field coupling function $f(\phi)$ and $\rho_m$ and indirectly as a function of Hubble parameter $H$ and $\rho_{tac}$. From (\ref{rhoD}) and (\ref{omega}) we find
\begin{equation}\label{rhot}
\dot\rho_{D} = -2H\rho_{D}\frac{\sqrt{\Omega_D}}{n}.
\end{equation}
From \ref{rhot} and (\ref{ct}), the EoS parameter of the original ADE in flat universe is
\begin{eqnarray}\label{omega1}
\omega_D&=&-1+\frac{2\sqrt{\Omega_D}}{3n}-\frac{Q}{3H\rho_D}.
\end{eqnarray}
From (\ref{omega}) and relation $\dot\Omega_D=\Omega'_DH$, we obtain
\begin{eqnarray}\label{omegap}
\Omega'_D &=& \Omega_D (-2\frac{\dot H}{H^2}-\frac{2\sqrt\Omega_D}{n}),
\end{eqnarray}
where the prime means derivative with respect to $x = \ln a$. Using the Friedman
equation (\ref{fried11}) and equations (\ref{cons}),  (\ref{rhoD}),  (\ref{omega}) and  (\ref{ct1}), one can
show that
\begin{eqnarray}\label{dH}
\frac{\dot H}{H^2}&=&-\frac{3}{2}(1+\Omega_D(-1+\frac{2\sqrt{\Omega_D}}{3n}-\frac{Q}{3H\rho_D})).
\end{eqnarray}
One then can rewrite the equation for original ADE as
\begin{eqnarray}\label{omegap1}
\Omega'_D &=& \Omega_D (1-\Omega_D)(3-\frac{2\sqrt{\Omega_D}}{n})-\frac{\Omega_m\Omega_D\dot f}{H}.
\end{eqnarray}
By using relation $\frac{d}{dx}=-(1+z)\frac{d}{dz}$ we can express $\Omega_D$ as
\begin{eqnarray}\label{omegaz}
\frac{d\Omega_D}{dz} &=& -(1+z)^{-1}(\Omega_D (1-\Omega_D)(3-\frac{2\sqrt{\Omega_D}}{n})-\frac{\Omega_m\Omega_D\dot f}{H}).
\end{eqnarray}
Now by using $\rho_{tac}$ with $\rho_D$ and relations $\rho_{tac}=\rho_D=3H^2M_p^2\Omega_D$ and $\omega_D=\dot\phi^2-1$, we find
\begin{eqnarray}
V(\phi) &=&\rho_{tac}\sqrt{1-\dot\phi^2}=3H^2M_p^2\Omega_D(1-\frac{2\sqrt\Omega_D}{3n}+\frac{Q}{3H\rho_D})^{1/2},\label{V}\\
\dot\phi&=&\sqrt{1+\omega_D}=(\frac{2\sqrt\Omega_D}{3n}-\frac{Q}{3H\rho_D})^{1/2}.\label{dphi}
\end{eqnarray}
Alternatively, the equation (\ref{dphi}) can rewritten as
\begin{eqnarray}\label{phip}
\phi' &=& H^{-1}(\frac{2\sqrt\Omega_D}{3n}-\frac{Q}{3H\rho_D})^{1/2},
\end{eqnarray}
or equivalently
 \begin{eqnarray}\label{phiz}
\frac{d\phi}{dz} &=&\frac{1}{H(1+z)}(\frac{2\sqrt\Omega_D}{3n}+\frac{Q}{3H\rho_D})^{1/2}.
\end{eqnarray}
Also by using Eq. (\ref{dH}) we can write
 \begin{eqnarray}\label{H}
\frac{dH}{dz} &=&-H(1+z)^{-1}(\frac{3\Omega_D}{2}-\frac{\Omega_D^{3/2}}{n}-\frac{3}{2}-\frac{\Omega_m\dot f}{2H}),
\end{eqnarray}
 where the sign is arbitrary and can be changes by a redefinition of the field, $\phi \rightarrow -\phi$. Then, by fixing the field amplitude at the present era to be zero, one can easily obtain the dynamic of the agegraphic tachyon field. It is difficult to solve equations (\ref{omegaz}),(\ref{phiz}) and (\ref{H}) analytically, however, the evolutionary form of the interacting agegraphic
tachyon field  $\phi$ and $\Omega_{tac}$ can be easily obtained integrating it numerically
from $z = 0$ to a given value $z$. In addition, from the constructed agegraphic
tachyon model, the evolution of $V(\phi)$ with respect to $\phi$ can be determined.
In the following, we assume that the function $f(\phi)$ behaves exponentially as $f(\phi)=f_0e^{b\phi(z)}$. Since $\dot{f}(\phi)$ is present in the interaction term $Q$, the parameters that determine the dynamics of the interaction are $b$ and $f_0$ together with $\rho_m$ and $\gamma$, the energy density and EoS parameter of the matter, respectively. Note that $f_0=0$ or $b=0$ leads to the absence of the interaction. We will do numerical calculation after the best fit analyzing of our model in section IV.

\section{TACHYON RECONSTRUCTION OF THE NEW ADE}
Next, we introduce the new ADE with the time scale as the conformal time $\eta$. The model benefits some new features that overcomes unsatisfactory aspects of original ADE. For instance, in the original ADE one can not represent the matter-dominated epoch while the new ADE can \cite{wei1}. The energy density of the new ADE is
\begin{eqnarray}
\rho_{D}&=&\frac{3n^2M_p^2}{\eta^2},\label{rho1}
\end{eqnarray}
with the conformal time $\eta$ as
\begin{equation}\label{T}
\eta=\int_0^a\frac{da}{Ha^2}.
\end{equation}
The density parameter of the new ADE is then
\begin{eqnarray}
\Omega_{D}&=&\frac{n^2}{H^2\eta^2}.\label{omega2}
\end{eqnarray}
From Eqs. (\ref{rho1}) and (\ref{omega2}) we obtain
\begin{eqnarray}
\dot\rho_{D}&=&-2H\frac{\sqrt\Omega_D}{na}\rho_D.\label{drho2}
\end{eqnarray}
Using (\ref{drho2}) and (\ref{ct}), the EoS parameter of the new ADE will be
\begin{eqnarray}\label{omegan}
\omega_D&=&-1+\frac{2\sqrt{\Omega_D}}{3na}-\frac{Q}{3H\rho_D}.
\end{eqnarray}
We also find the equation of motion for density parameter in new ADE as
\begin{eqnarray}\label{omegap2}
\Omega'_D &=& \Omega_D (1-\Omega_D)(3-\frac{2\sqrt{\Omega_D}}{na})-\frac{\Omega_m\Omega_D\dot f}{H},
\end{eqnarray}
or equivalently
\begin{eqnarray}\label{omegazn}
\frac{d\Omega_D}{dz} &=& -(1+z)^{-1}(\Omega_D (1-\Omega_D)(3-\frac{2\sqrt{\Omega_D}}{na})-\frac{\Omega_m\Omega_D\dot f}{H}).
\end{eqnarray}
Now, by using Eqs. (\ref{omega2}) and (\ref{omegan}) we obtain the tachyon potential
and derivative of the scalar field as
\begin{eqnarray}
V(\phi) &=&3H^2M_p^2\Omega_D(1-\frac{2\sqrt\Omega_D}{3na}+\frac{Q}{3H\rho_D})^{1/2},\label{V2}\\
\dot\phi&=&(\frac{2\sqrt\Omega_D}{3na}-\frac{Q}{3H\rho_D})^{1/2}.\label{dphin}
\end{eqnarray}
Equation (\ref{dphin}) as
\begin{eqnarray}\label{phipn}
\phi' &=& H^{-1}(\frac{2\sqrt\Omega_D}{3na}-\frac{Q}{3H\rho_D})^{1/2},
\end{eqnarray}
or
 \begin{eqnarray}\label{phizn}
\frac{d\phi}{dz} &=&\frac{1}{H(1+z)}(\frac{2\sqrt\Omega_D}{3na}+\frac{Q}{3H\rho_D})^{1/2}.
\end{eqnarray}
Similarly for new ADE we find
 \begin{eqnarray}\label{Hn}
\frac{dH}{dz} &=&-H(1+z)^{-1}(\frac{3\Omega_D}{2}-\frac{\Omega_D^{3/2}}{na}-\frac{3}{2}-\frac{\Omega_m\dot f}{2H}).
\end{eqnarray}
Again, it is difficult to solve Eqs. (\ref{omegazn}),(\ref{phizn}) and
(\ref{Hn}) analytically, however, the evolutionary form of the interacting new agegraphic tachyon field
$\phi$ and $\Omega_{tac}$ can be easily obtained by integrating it numerically from $z = 0$ to a given value
$z$. In addition, from the constructed agegraphic tachyon model, the evolution of $V(\phi)$
with respect to $\phi$ can be determined.

\section{Observational best fitting with Hubble parameter, $H(z)$}

Now, we study the constraints on the model parameters using $\chi^2$ method, utilizing recent observational data, including the Hubble parameter as a function of the redshift, the baryonic acoustic oscillation (BAO) distance ratio and the cosmic microwave background (CMB) radiation.

We solve the set of coupled nonlinear partial differential equations, (\ref{omegaz}), (\ref{phiz}), (\ref{H}). For best fitting the model for the parameters $n$ ,$b$ and the initial conditions $\Omega_D(0)$ and $H(0)$ with the most recent observational data for Hubble parameter, we employe the $\chi^2$ method. We constrain the parameters including the initial conditions by minimizing the $\chi^2$ function given as
\begin{widetext}
\begin{equation}\label{chi2}
    \chi^2_{Hub}(n ,b; \Omega_D(0), H(0))=\sum_{i=1}^{14}\frac{[H^{th}(z_i|n ,b; \Omega_D(0), H(0)) - H^{obs}(z_i)]^2}{\sigma_{Hub}^2(z_i)},
\end{equation}
\end{widetext}
where the sum is over the cosmological dataset. In (\ref{chi2}),  $H^{th}$ and $H^{obs}$ are the Hubble parameters obtained from the theoretical model and from observation, respectively. Also, $\sigma_{Hub}$ is the estimated error of the $H^{obs}$ where obtained from observation \cite{Cao}.

We also constraint the model parameters by using CMB. In CMB analysis, the shift parameter $R$ \cite{r1}-\cite{r2} where comprises the necessary observational information from CMB, constraints the model parameters by minimizing
\begin{equation}\label{chicmb}
    \chi^2_{CMB}=\frac{[R-R_{obs}]^2}{\sigma_R^2},
\end{equation}
where $R_{obs} = 1.725\pm0.018$ \cite{r3}, is taken WMAP7. The shift parameter theoretically is defined by defined as
\begin{equation}\label{r}
    R\equiv\Omega_{m0}^{1/2}\int_0^{z_{CMB}}\frac{dz'}{E(z')},
\end{equation}
where $z_{CMB} = 1091.3$.
Furthermore, we  use the BAO data to constraint the model. By applying the 2dF Galaxy Redsihft Survey and SDSS data, we find the BAO distance ratio at $z = 0.20$ and $z = 0.35$ \cite{snss}-\cite{He}. The distance ratio is
\begin{equation}\label{drbao}
   \frac{D_V(z=0.35)}{D_V(z=0.20)}=1.736\pm0.065,
\end{equation}
where $D_V(z)$ given by
\begin{equation}\label{bao}
    D_V(z_{BAO})=[\frac{z_{BAO}}{H(z_{BAO})}(\int_0^{z_{BAO}}\frac{dz}{H(z)})^2]^{1/3}.
\end{equation}
therefore, by calculating the following $\chi^2$ statistics
\begin{equation}\label{chibao}
    \chi^2_{BAO}=\frac{[(D_V(z=0.35)/D_V(z=0.20))-1.736]^2}{0.065^2}\cdot
\end{equation}
one can find new constraints on model parameters. The constraints from a combination of $H(z)$ Ia, BAO and CMB can be obtained by minimizing $\chi^2_{Sne}+\chi^2_{BAO}+\chi^2_{H(z)}$ in two original and new ADE scenarios. The results are shown in tables \ref{table:1} and \ref{table:2}.

\begin{table}[ht]
\caption{best fit values of original ADE model} 
\centering 
\begin{tabular}{c|c|c|c} 
\hline\hline 
observational data  &  H(z) \ & H(z)+CMB \ & H(z)+CMB+BAO \\ [4ex] 
\hline 
H(0) (Km/s/Mpc) & 71 & 71 & 72 \\ 
\hline 
$\Omega_D$(0) & 0.72 & 0.73 & 0.73 \\
\hline 
n & 18.6 & 7.2 & 22.2\\
\hline 
b & -0.7 & 0.2 & -0.4 \\
\hline
\end{tabular}
\label{table:1} 
\end{table}\

\begin{table}[ht]
\caption{best fit values of NADE model} 
\centering 
\begin{tabular}{c|c|c|c} 
\hline\hline 
observational data  &  H(z) \ & H(z)+CMB \ & H(z)+CMB+BAO \\ [4ex] 
\hline 
H(0) (Km/s/Mpc) & 71 & 71 & 72 \\ 
\hline 
$\Omega_D$(0) & 0.72 & 0.73 & 0.73 \\
\hline 
n & 24 & 8.3 & 38.3\\
\hline 
b & -0.9 & 0.3 & -0.6 \\
\hline
\end{tabular}
\label{table:2} 
\end{table}\

\begin{figure}
\centering
\includegraphics[width=0.4\textwidth]{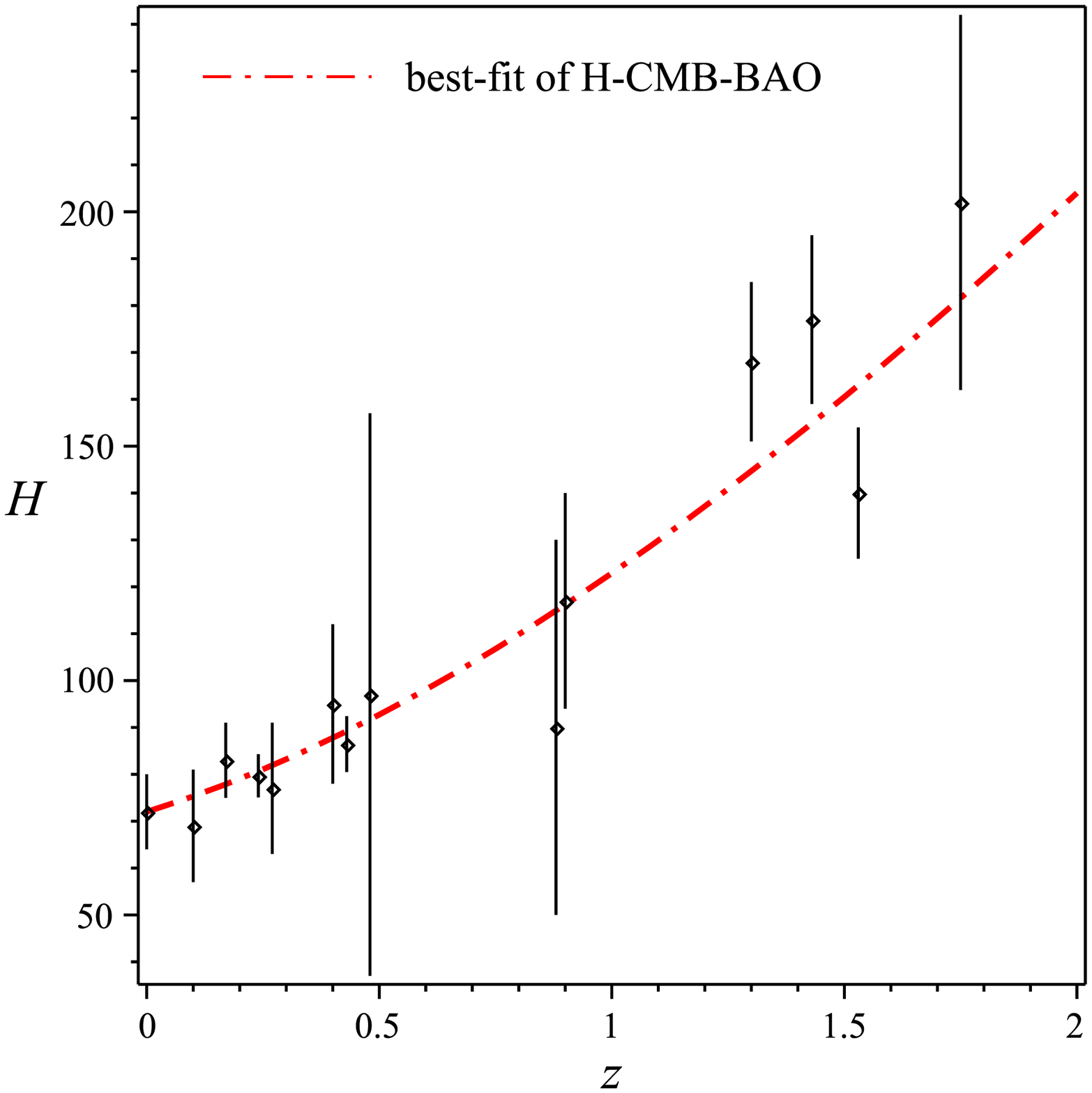}
\includegraphics[width=0.4\textwidth]{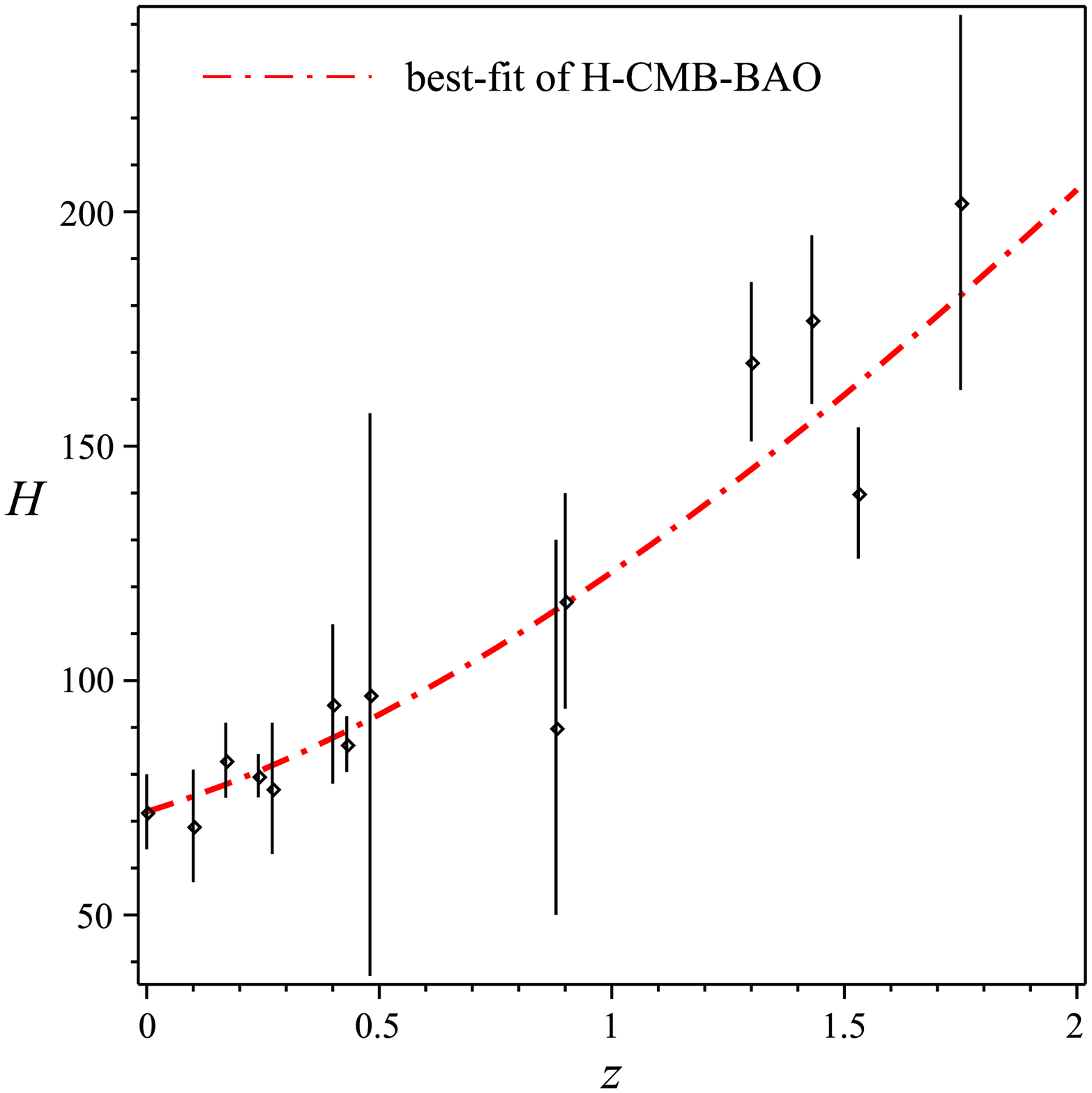}
\caption{ the best fitted H(z)+CMB+BAO, for the original and new ADE model}\label{fig:1}
\end{figure}

From the Tables \ref{table:1} and \ref{table:2}, and Fig:\ref{fig:1} the results show that the initial conditions for dynamical variables $\Omega_D$ and $H$ are not sensitive to the cosmological models of original and new ADE. They are not also sensitive to the CMB and BAO data. However, the model parameters $n$ and $b$ are very dependent on both cosmological models and observational dataset.

\section{Cosmological test}

\begin{figure}
\centering
\includegraphics[width=0.4\textwidth]{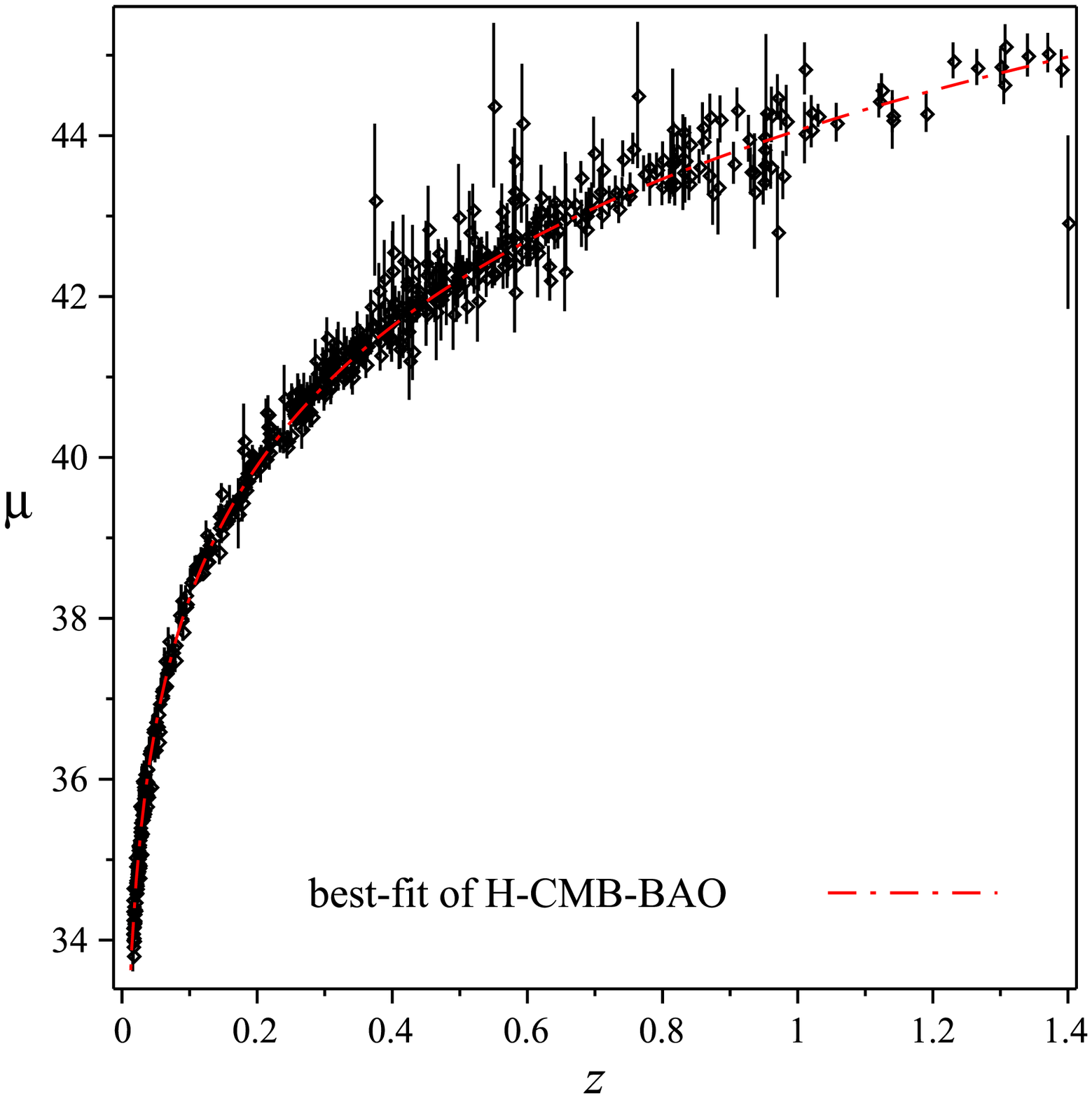}
\includegraphics[width=0.4\textwidth]{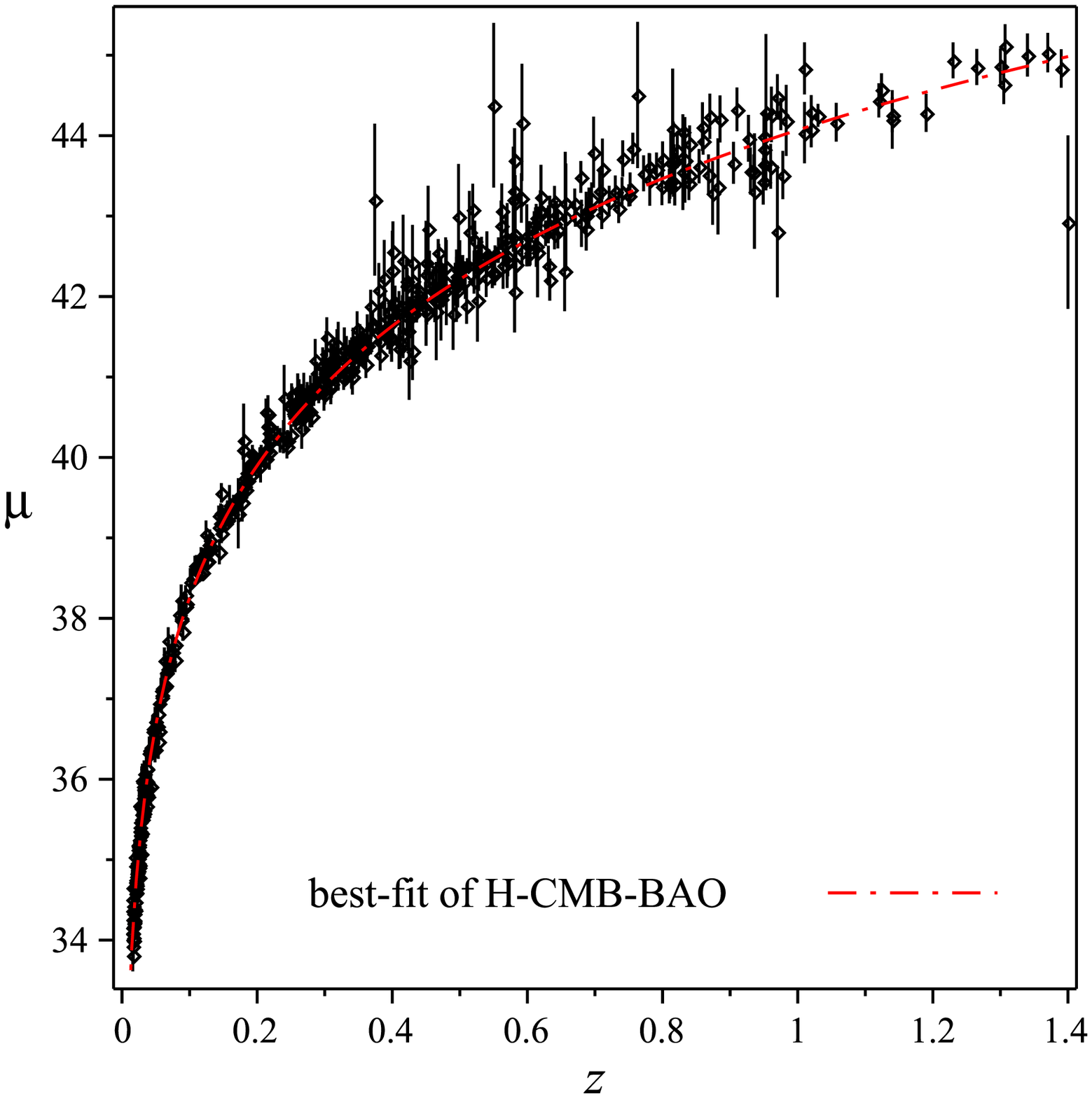}
\caption{The best fitted distance modulus for the original and new ADE model}\label{fig2}
\end{figure}

We have already best fitted our model with the current observational data for Hubble parameter. Now, we test our model against recent observational data for the best fitted distance modulus in both original and new ADE, as shown in Fig:\ref{fig2}. The graphs show that the model is in good agreement  with the observational data.

\begin{figure}
\centering
\includegraphics[width=0.4\textwidth]{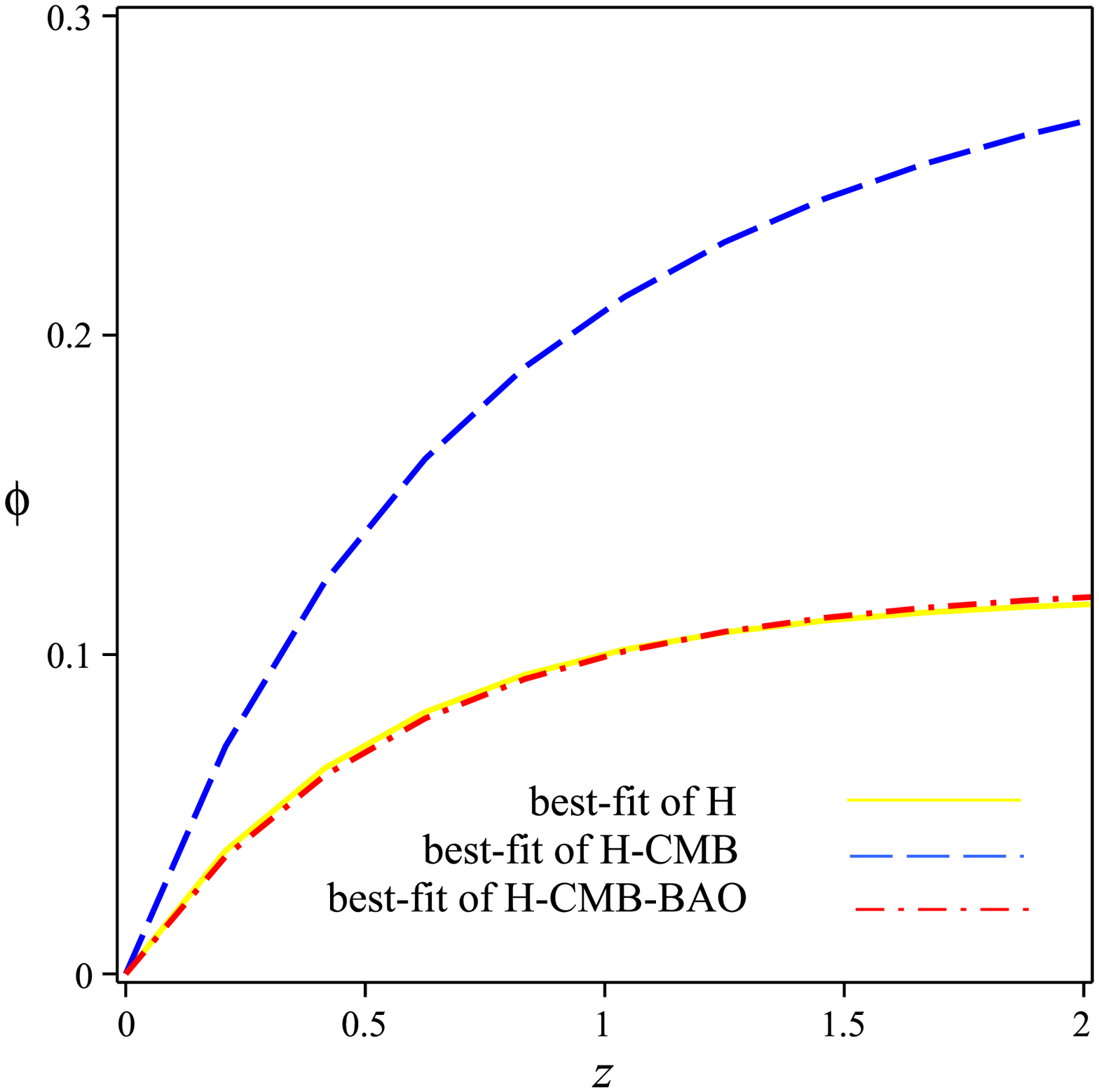}
\includegraphics[width=0.4\textwidth]{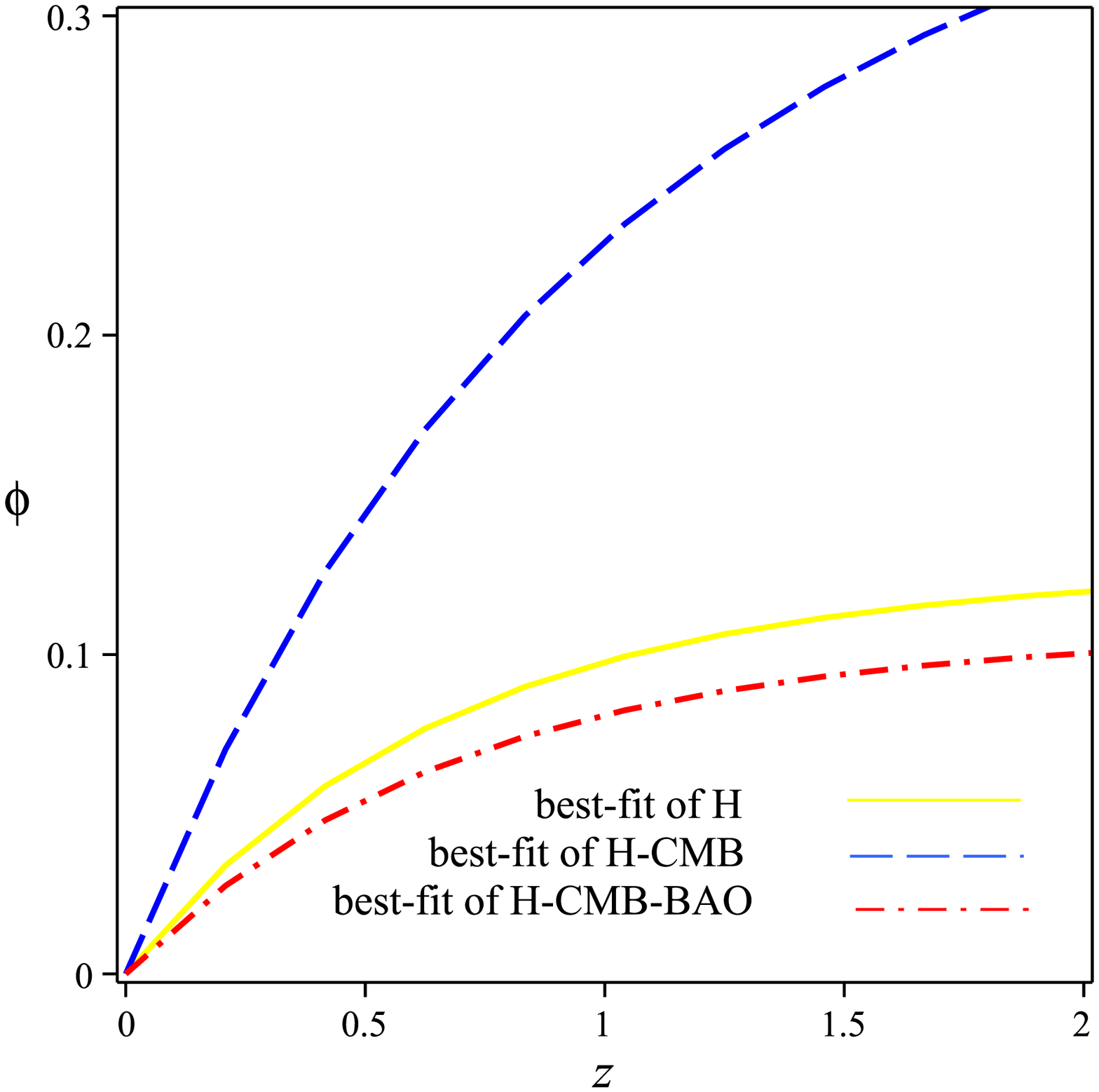}
\caption{yellow(solid), blue(dash) and red(dash-dot) lines show the best fitted $\phi(z)$ , respectively, for the original and new ADE model}\label{fig3}
\end{figure}

\begin{figure}
\centering
\includegraphics[width=0.4\textwidth]{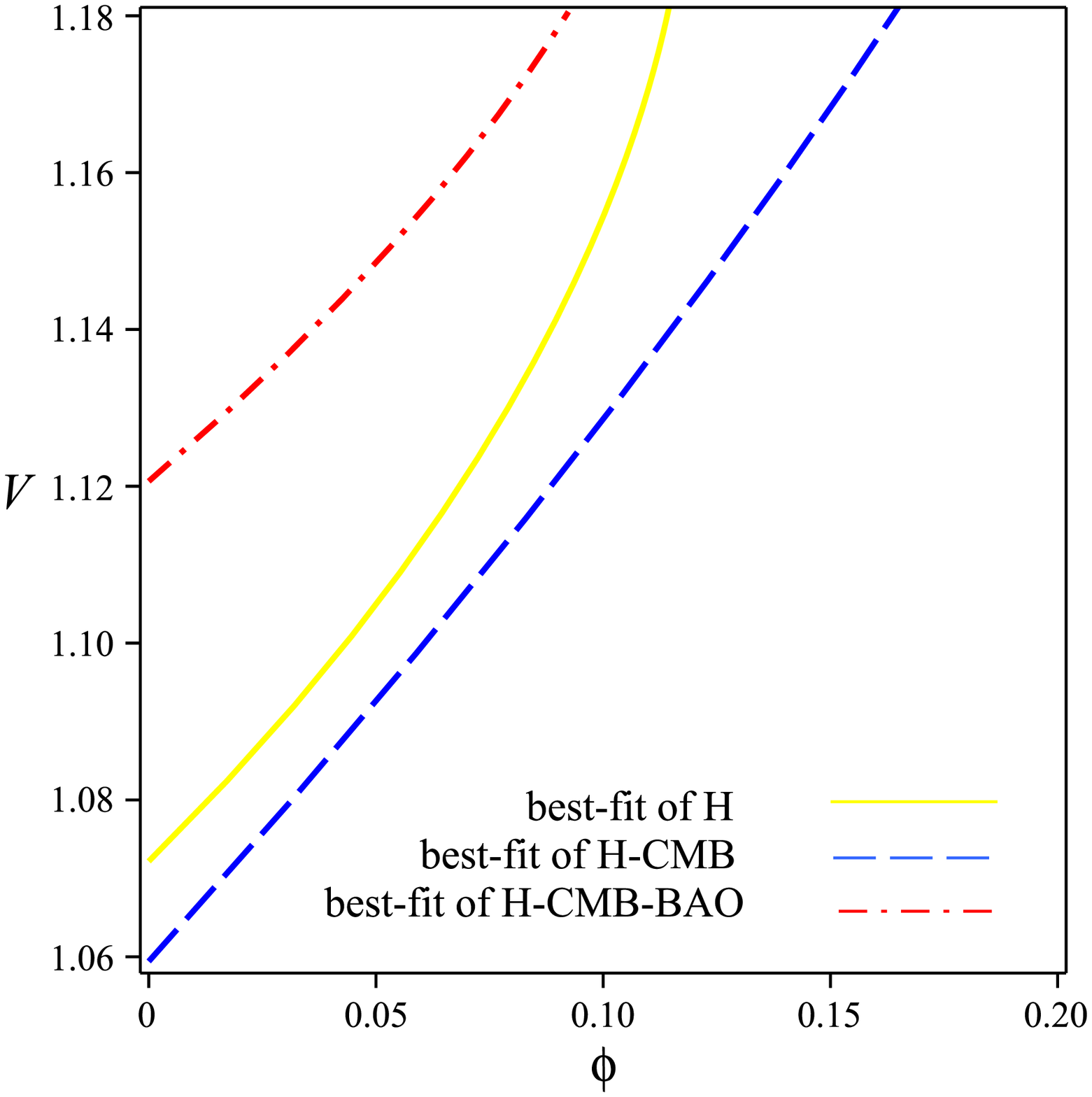}
\includegraphics[width=0.4\textwidth]{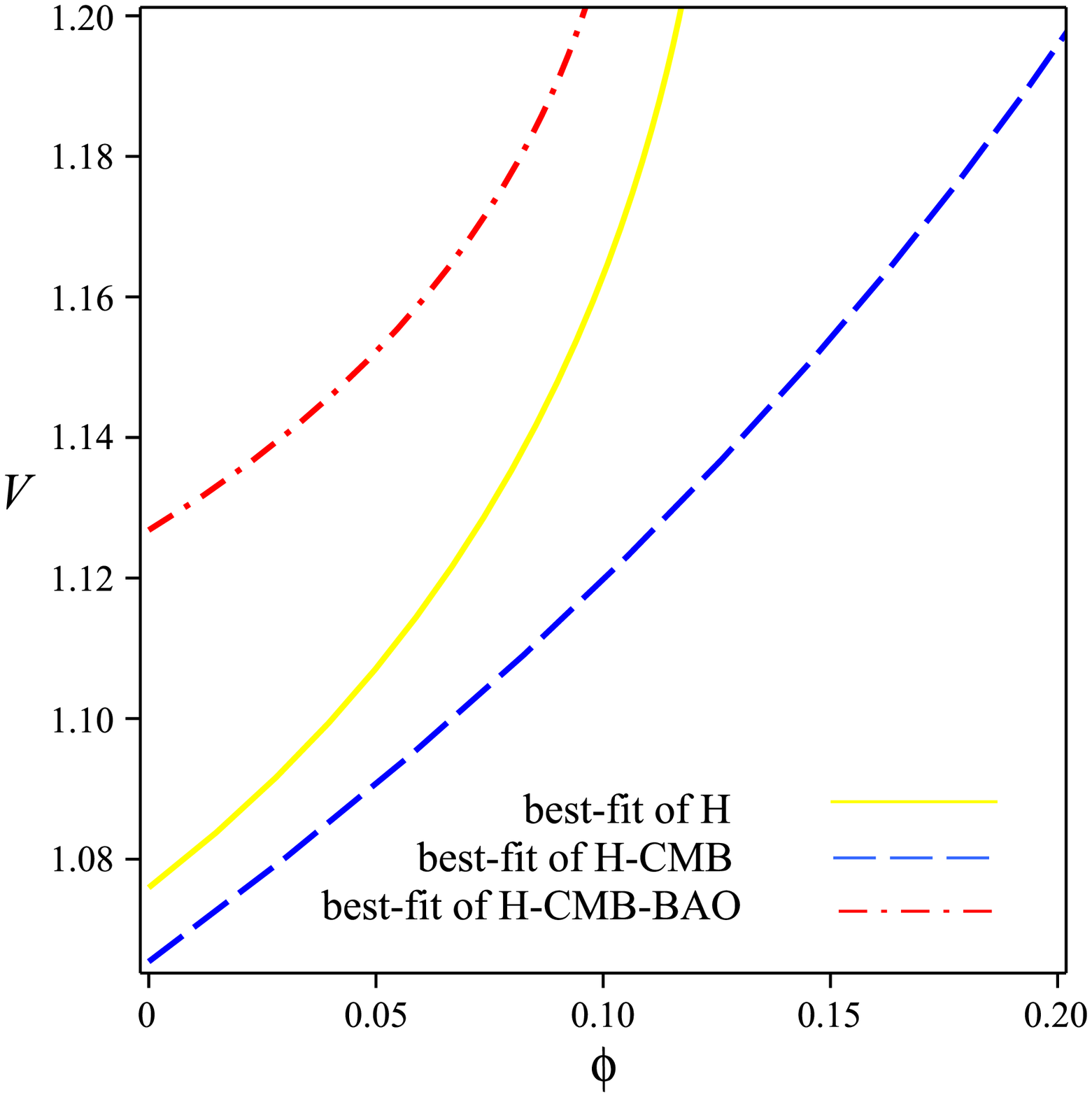}
\caption{yellow(solid), blue(dash) and red(dash-dot) lines show the best fitted $V(\phi)$, respectively, for the original and new ADE model}\label{fig4}
\end{figure}

In addition for both models we plotted the reconstructed potential and scalar field $V(\phi)$ and $\phi(z)$ respectively using the best fitted model parameters (see Figs.\ref{fig3} and \ref{fig4}).

In Fig.\ref{fig3}, we see that the dynamics of the best fitted and reconstructed scalar field depends on the observational data and also cosmological models. The graphs show a decreasing trend with decreasing redshift. Similar behavior can be seen in the graphs for the reconstructed potential function with the best fitted model parameters.

\begin{figure}
\centering
\includegraphics[width=0.4\textwidth]{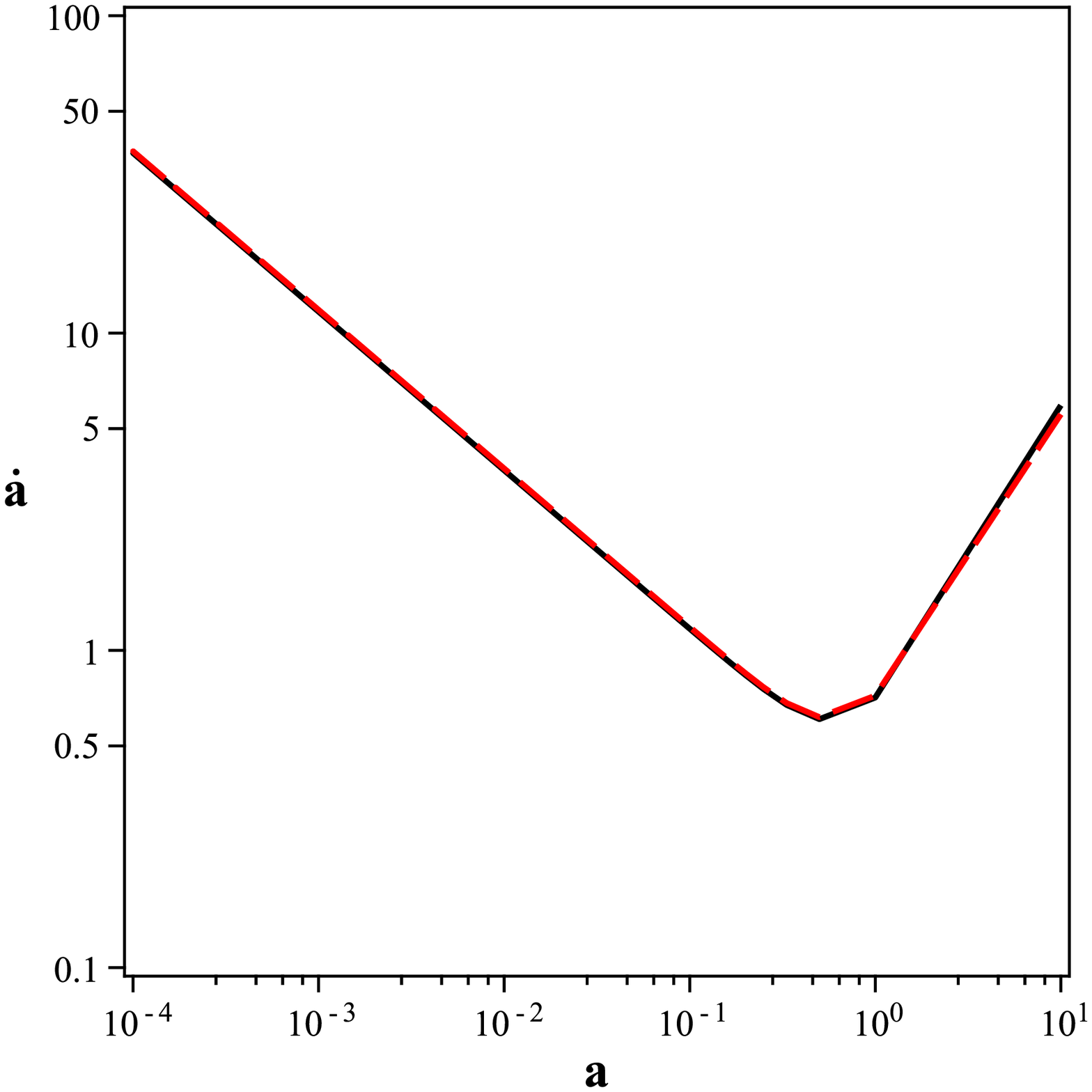}
\caption{Phase space of the scale factor for $\Omega_D = 0.74$ and the best fitted model parameters. At present epoch, we choose $a = 1; z = 0$. The transition from  decelerated expansion to accelerated one is shown in both old (dashed red) and new (solid black) ADE scenarios. }\label{fig5}
\end{figure}

In \cite{Baglaa2}, based on different forms of tachyon potential, the behavior of the scale factor and the duration of the accelerated expansion of the universe is discussed. Here, in Fig.\ref{fig5}, we have plotted the phase portrait for the scale factor in both old and new ADE scenarios. The graph shows that the current tachyon dominated universe creates late time acceleration and earlier dark matter domination produces the deceleration phase in te past. It also illustrates the dynamics of cosmic scale factor in both cases of old and new ADE.

\begin{figure}
\centering
\includegraphics[width=0.4\textwidth]{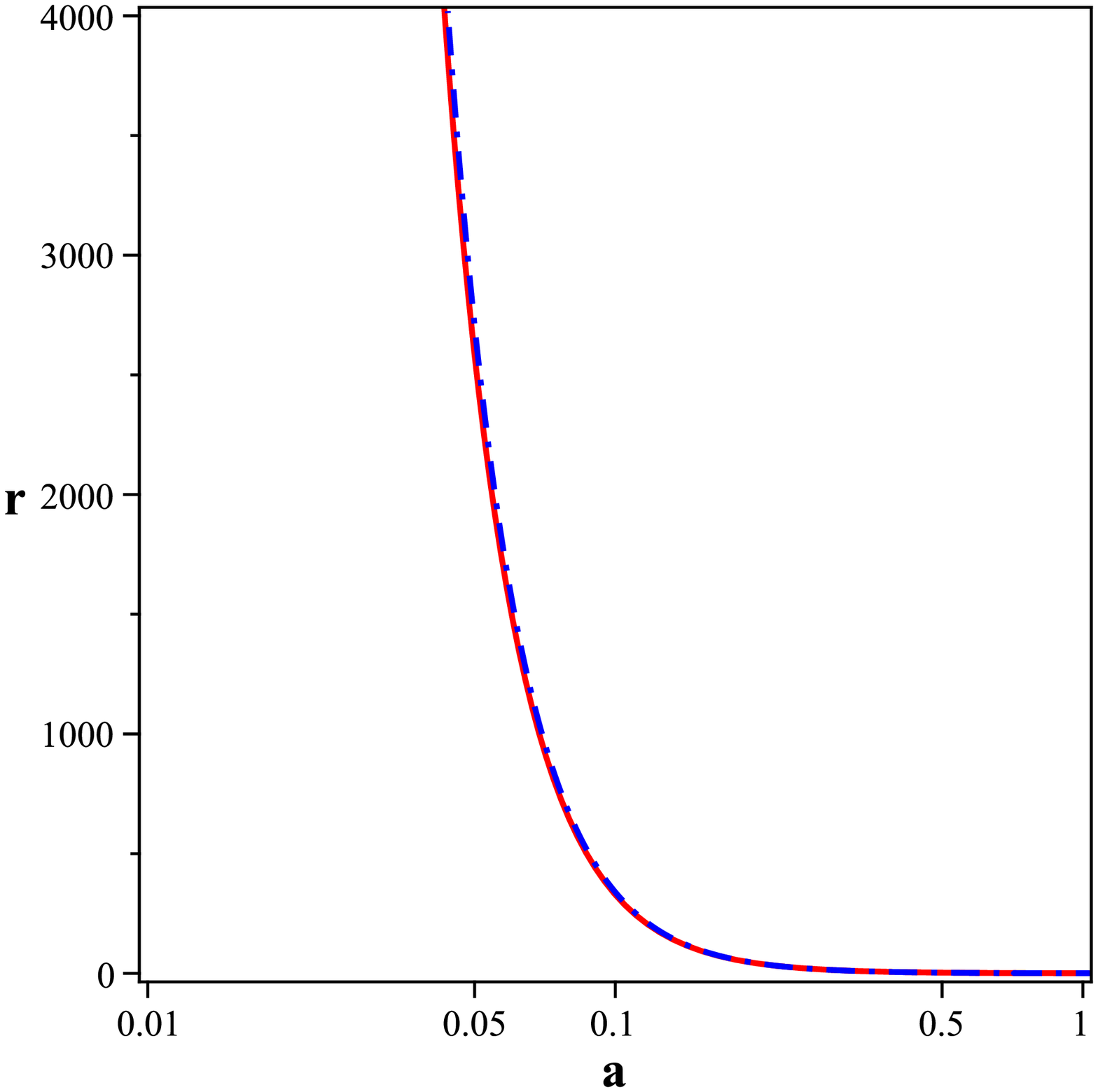}
\caption{The behaviors of the ratio $r = \rho_c/\rho_d$ in two cases of old and new ADE scenarios.}
\label{fig6}
\end{figure}

Using the fitting result, we have also studied the coincidence problem. The ratio of energy densities between DM and DE, $r = \rho_{DM}/\rho_D$, and its evolution is plotted with respect to the scale factor in \ref{fig6}. From the graph we observe a slower change of $r$ in the current epoch of the universe expansion in both cases of old and new ADE scenarios. Also, the ratio is about one to one in the late time era. In comparison to $\Lambda CDM$ model, the period when energy densities of DE and DM are comparable is longer due to the interaction between DE and DM, see Fig.\ref{fig6}. This in turn ameliorates the coincidence problem. Similar to the phase space of the scale factor, the ratio of energy densities is not affected by the new ADE model.

\section{Summary}

In this paper, we investigate the original and new interacting agegraphic
dark energy models in tachyon cosmology. We assume that the matter field act as DM and the tachyon filed plays the role of original or new ADE. We also assume that these two dark components interact and the interaction term emerges from the model rather being inserted into the formalism as an external source. We first constraint the model parameters and initial conditions with the observational data for hubble parameter using chi-squared statistical method. The result shows that the initial conditions are insensitive to the cosmological models and observational data whereas the model parameters highly depend on them. Then based on the best fitted evolutionary behavior of the interacting original and new ADE, we test the model against observational data for distance modulus and also reconstruct the tachyon potential and scalar field. Further, we compute the scale factor velocity and the ratio of dark sectors with respect to the scale factor of the universe in both scenarios and revisit the coincidence problem.

\end{document}